# Materials Relevant to Realizing a Field-Effect Transistor based on Spin-Orbit Torques

Phillip Dang, Zexuan Zhang, Joseph Casamento, Xiang Li, Jashan Singhal, Darrell G. Schlom, Daniel C. Ralph, Huili Grace Xing, *Senior Member, IEEE* and Debdeep Jena, *Senior Member, IEEE*

*Abstract*—**Spin-orbit torque is a promising mechanism for writing magnetic memories, while field-effect transistors are the gold-standard device for logic operation. The spin-orbit torque field effect transistor (SOTFET) is a proposed device that couples a spin-orbit-torque-controlled ferromagnet to a semiconducting transistor channel via the transduction in a magnetoelectric multiferroic. This allows the SOTFET to operate as both a memory and a logic device, but its realization depends on the choice of appropriate materials. In this report, we discuss and parametrize the types of materials that can lead to a SOTFET heterostructure.**

## I. INTRODUCTION

Though charge-based memory has made rapid advances in recent years via the NAND Flash and its recent 3D vertical incarnation, they fall short in comparison with magnetic memories in the metrics of reading/writing energies and durability [1], [2]. The competitiveness of modern magnetic memories began with discovery of giant magnetoresistance (GMR) of magnetic multilayers in 1988 [3], which introduced a method to utilize the spin of an electron to detect the magnetization orientation of a ferromagnet and enabled an efficient reading mechanism [4]. In the following years, advances in write operations for GMR-based memories were discovered via the experimental observations [5], [6] of theoretically predicted spin-transfer torque (STT) [7] and that of spin-orbit torque (SOT) [8]-[10]. Today, the STT and SOT mechanisms are under investigation because of the lower energy consumption they offer compared to previous GMR-based memories. More importantly, because of the fast read/write times achieved in STT and SOT switching, magnetic memories are no longer limited solely to permanent storage; they are currently being explored for their capacity for nonvolatile, infinite endurance, energy efficient, high density magnetoresistive random access memories (MRAM) [11], [12].

Despite paving the way for high density magnetic memories, the GMR and tunneling magnetoresistance (TMR) mechanisms produce a limited change in resistance between the distinct memory states of the magnetic layers resulting in longer read times, higher read energies, and potential read errors in MRAM. By comparison, the change in resistance between the on and off states of a semiconductor field-effect transistor (FET) is much larger, by several orders of magnitude; a magnetic memory could benefit significantly if had access to the large change in resistance exhibited by a semiconductor. The large change in resistance not only increases the speed, energy efficiency, and reliability of reading magnetic memories but also opens the possibility of using magnetic memory devices in logic circuits, which require the large on/off ratios offered by FETs. This could result in new architectures, such as processing in memory, that would significantly improve computational speed as a whole. One way to achieve this device benefit would be to exploit the newly developed family of multiferroics – materials that are simultaneously magnetically ordered and ferroelectric – to sense the change in spin orientations from a SOT-ferromagnetic layer using exchange coupling to its magnetization on one side, and transmit this signal through its electric polarization to a semiconductor channel on the other. For this to happen, the magnetic order parameter should be coupled to the ferroelectric order parameter, which is the defining feature of a magnetoelectric multiferroic material. Such a device is called the SOTFET: the spin-orbit torque field-effect transistor. The SOTFET was recently proposed and analyzed by Li et al [13], and its device and circuit implications were analyzed [14]. In this paper, we focus on the choices of materials that are desirable to realize the SOTFET and their present level of maturity.

## II. THE SPIN-ORBIT TORQUE FIELD-EFFECT TRANSISTOR

The SOTFET consists of a semiconducting channel and a multiferroic/ferromagnet/spin-orbit material gate stack. Fig. 1a shows a cartoon of the SOTFET and possible current, magnetization, and electrical polarization directions that would constitute the "on" and "off" states of the transistor. We note that the actual directions of these vectors in a fully realized device could be different, depending on the ultimate choice of materials. Fig. 1b shows potential material choices for each layer of the SOTFET. From the top of the SOTFET gate stack to the channel, the SOTFET operates by using SOT to switch a ferromagnetic layer that is coupled to a magnetoelectric multiferroic. When the direction of the spontaneous magnetization of the magnetoelectric multiferroic is switched, the direction of its spontaneous electrical polarization is also deterministically switched. This change in polarization induces a change in electric field in the semiconducting channel that enhances or depletes the number of carriers in the semiconducting channel.

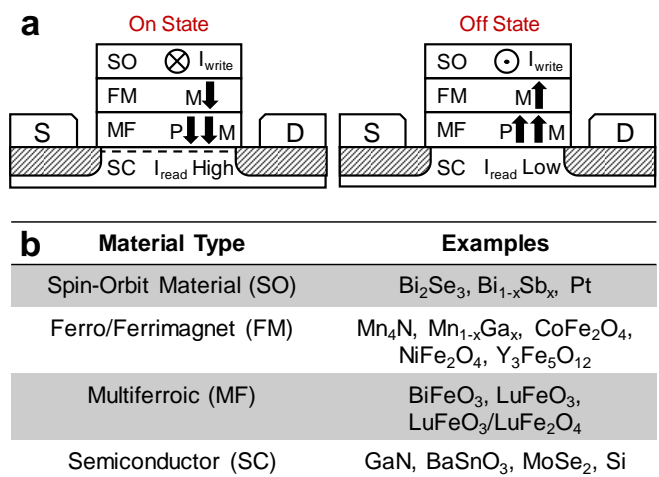

Fig. 1. (a) Cartoon of SOTFET device for potential on and off states. The arrows represent possible directions for magnetization (M) and electrical polarization (P) (b) Table of possible materials for implementation in the SOTFET.

| Material Type | Examples |
|---|---|
| Spin-Orbit Material (SO) | $Bi_2Se_3$, $Bi_{1-x}Sb_x$, Pt |
| Ferro/Ferrimagnet (FM) | $Mn_4N$, $Mn_{1-x}Ga_x$, $CoFe_2O_4$, $NiFe_2O_4$, $Y_3Fe_5O_{12}$ |
| Multiferroic (MF) | $BiFeO_3$, $LuFeO_3$, $LuFeO_3/LuFe_2O_4$ |
| Semiconductor (SC) | GaN, $BaSnO_3$, $MoSe_2$, Si |

### A. Spin-Orbit Torques

Magnetic memory has typically been based on the switching of the magnetic orientation of a ferromagnet and presents a non-volatile and high-endurance form of information storage. In order to electrically detect the magnetic orientation of the ferromagnet, spin valves and magnetic tunnel junctions, based on GMR and TMR, have been widely researched. In these devices, a change in resistance up to ~600% [15] can occur when the magnetization of a ferromagnet is switched. Spin-transfer torque (STT) is a phenomenon where nanoscale magnets can be switched by transferring the spin angular momentum of electrons from a spin-polarized current, and the maximum angular momentum transferred by one electron in $\hbar/2$ [11]. Spin-orbit torque (SOT) takes this idea of transferring spin angular momentum but utilizes a pure spin current generated within a spin-orbit (SO) material [12]. For example, when a charge current flows in the SO material shown in Fig. 1a and 1b, it can generate a transverse-flowing spin current that can transfer non-equilibrium spin angular momentum to an adjacent ferromagnet (FM). Each electron in the current flowing adjacent to the ferromagnet in a SOT device can transfer angular momentum several times, leading to a more efficient magnetic switch than in STT [12]. The SOTFET uses SOT as the switching mechanism, as illustrated in Fig. 1a, but presents an alternative device structure to GMR- and TMR-based devices for the readout.

### B. Ferromagnet/Ferroelectric Coupling

The SOTFET requires that the switching of a ferromagnet by SOT results in the deterministic switching of the spontaneous electrical polarization in a multiferroic. In a magnetoelectric multiferroic, the electrical polarization and magnetization are strongly coupled. For sufficient coupling between ferromagnetism and ferroelectricity and a sufficiently low energy barrier for switching the electrical polarization, it is expected that switching the spontaneous magnetization of the multiferroic will deterministically switch the spontaneous electrical polarization of the multiferroic. Furthermore, the magnetization of the ferromagnetic layer must be exchange-coupled to that of the multiferroic layer. Such exchange coupling has been demonstrated most notably in a $BiFeO_3$-$Co_{0.9}Fe_{0.1}$ heterostructure with electric field control of magnetism [16], but the converse effect, which would enable the operation of the SOTFET, has not been demonstrated yet in this material. This is may be due to the trend in multiferroics, the interactions that govern ferroelectricity are much larger in magnitude than those that govern magnetic ordering [17].

### C. Gating of the Field-Effect Transistor

In a FET, an electric field applied through a voltage on the gate determines the density of charge carriers that conduct current in the semiconducting channel. In the SOTFET, the electrical polarization of the multiferroic supplies the electric field required to gate the transistor. In the ideal situation, the change in electric field when the multiferroic switches would be sufficient to switch the FET from its on state to its off state, causing a large change in the semiconductor channel resistance. This on/off current ratio of the FET is determined by (1) properties of the semiconductor, such as the bandgap, carrier density, and carrier mobility, as well as (2) the extent to which carrier concentration can be controlled by the electric field produced by the multiferroic, which is determined by net polarization-charge density, interface density of states, and electrostatic design of the channel (e.g. doping, layer thickness, geometry, etc.). The desired property for SOTFET operation is when the saturation polarization of the multiferroic is comparable to the on-state mobile charge sheet density in the semiconducting channel, which is typically in the $10^{12}$-$10^{13}$/cm$^2$ regime and corresponds to a charge of < 2 μC/cm$^2$.

## III. SEMICONDUCTORS AND MULTIFERROICS

The semiconductor material desirable for the SOTFET should be able to intrinsically exhibit a large on/off ratio of the conducting channel in response to the field effect, and possess a high carrier mobility to reduce ohmic losses. That by itself is achievable in a large range of semiconductors for which FETs exist. The SOTFET however needs the field effect in the semiconductor to be via the switchable electric polarization of a multiferroic layer with which it is intimately integrated. The multiferroic layer should ideally have a sufficient band offset with the semiconductor band of interest and should be of a sufficient energy bandgap to avoid electrical shorting or ambipolar injection. Because of the limited choices of suitable multiferroics, the choice of semiconductors for the device are driven by their potential for integration with the suitable multiferroic.

Magnetoelectric multiferroics are a class of materials that couple ferroic order parameters, namely ferromagnetism and ferroelectricity. For many electronic device applications, including the SOTFET, this coupling should persist above room temperature. This is shown in the shaded gray box in Fig. 2 as magnetic and ferroelectric transition temperatures should ideally be above 400 K. Magnetic order generally decreases as temperature increases due to thermal fluctuations; similarly, ferroelectrics undergo charge order fluctuations or phase transitions that give rise to a paraelectric state at higher temperatures. For these reasons, as well as the complexity of interactions that govern magnetoelectric coupling, it is difficult to obtain room temperature or near-room temperature magnetoelectric multiferroic materials. Only two magnetoelectric multiferroics have been demonstrated to deterministically switch near room temperature: BiFeO₃ [28] and LuFeO₃/LuFe₂O₄ superlattices [21]. Of these two, BiFeO₃ is the most widely studied as BiFeO₃ has the advantage of being thermodynamically stable, whereas LuFeO₃/LuFe₂O₄ superlattices are new artificial materials that are just beginning to be studied [29]. The family of relevant multiferroics is expected to expand in the future.

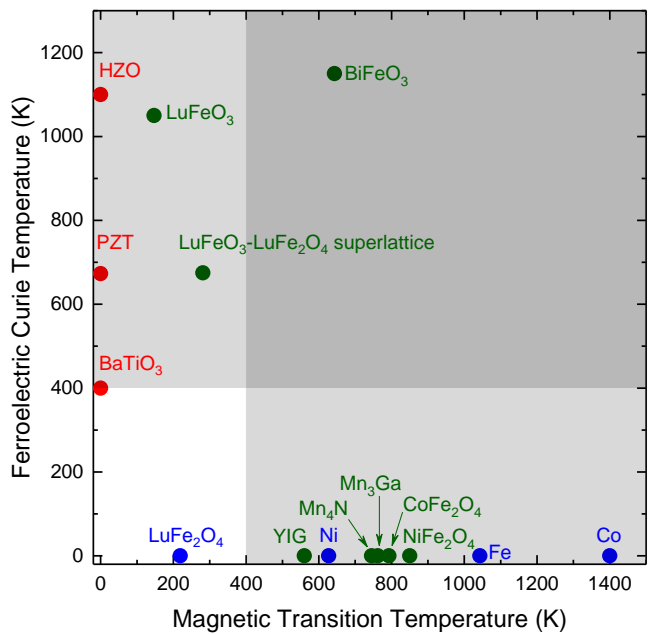

Fig. 2. Neel and Curie Temperatures of ferroelectrics [18]-[20] (red); ferromagnets [21] (blue); and ferrimagnets [21], [22]-[27] (green). Shaded regions represent transition temperatures above 400 K, which are desirable for applications.

It is feasible to integrate layered, 2D semiconductors on top of the multiferroic as well, instead of growing the multiferroic on a semiconductor. This geometry is an "inverted" SOTFET when compared to Fig 1. Though 2D or layered semiconductors in principle can be mechanically integrated with the multiferroic in a facile manner, the control of the chemical and electronic properties of the multiferroic/2D semiconductor interface will determine its actual usefulness in the SOTFET. Since this usefulness is currently unknown, however, we limit our discussion of compatible semiconductors to 3D semiconductors with promising compatibility with BiFeO₃ and LuFeO₃/LuFe₂O₄ superlattices.

### A. Multiferroic BiFeO₃

BiFeO₃ has primarily been grown epitaxially on insulating or metallic substrates to form capacitor structures for studying its multiferroic properties. Early efforts to integrate BiFeO₃ with the wide-bandgap semiconductor GaN resulted in a

(0001)∥(0001) epitaxial relationship, but required the use of SrTiO₃ and TiO₂ buffer layers [30]. From the point of view of epitaxial growth, the oxide semiconductor BaSnO₃ provides the possibility of a high mobility channel that is structurally matched with BiFeO₃. BiFeO₃ and BaSnO₃ are chemical compatible (both are oxides) and share the same crystal structure; both are perovskites. Importantly, BaSnO₃ is the highest electron mobility oxide perovskite known to date, with bulk room-temperature mobility exceeding 300 cm²/V·s [31]. This is because unlike perovskites such as SrTiO₃ where the localized d-orbitals of the Ti atom form the conduction band, the high mobility of BaSnO₃ derives from the delocalized s-orbitals of the Sn atoms. Coupled with the large bandgap (~3 eV) [32], and the potential to sustain large electric fields, it is an attractive semiconductor substrate for integration with BiFeO₃. Nevertheless, its band offsets with BiFeO₃ remain unexplored to date. Since the bandgap of BiFeO₃ (Eg~2.7 eV) [33] is only slightly smaller than that of BaSnO₃, it is important that the band offset be measured. It may be necessary to use a wider bandgap interlayers, for example thin layers in which Sr is alloyed with BaSnO₃ as the alloy (BaₓSr₁₋ₓ)SnO₃.

### B. Multiferroic LuFeO₃ and LuFeO₃/LuFe₂O₄ Superlattices

The main ingredient of the magnetoelectric multiferroic LuFeO₃/LuFe₂O₄ superlattices [21] is LuFeO₃. At room temperature it is ferroelectric and on cooling below 147 K becomes also multiferroic [34]. By integrating LuFeO₃ into the LuFeO₃-LuFe₂O₄ superlattice structure, the magnetic transition temperature raises to approximately 281 K. The LuFeO₃ polymorph of interest is hexagonal and metastable. There is also reason to believe that hexagonal LuFeO₃ may offer deterministic switching between **P** and **M** in its multiferroic state [35]. In contrast, the stable polymorph of LuFeO₃ is centrosymmetric thus neither ferroelectric nor multiferroic. The desired metastable polymorph of LuFeO₃ has been grown primarily on YSZ (111) substrates [36]. Nonetheless, hexagonal GaN or AlN substrates with c-plane orientation provide the correct symmetry matching [37]. There is a significant lattice mismatch, however, which would lead to defect formation in the LuFeO₃. The energy bandgap of LuFeO₃ is also rather small (~1 eV) [38], which implies a low chance of the desired large band offset to the underlying semiconductor channel, e.g., GaN or AlN; this band offset has not yet been measured. Nevertheless, the nitride semiconductor platforms offers a choice of mature and tested heterostructure platforms such as Al(Ga)N/GaN heterostructures, in which the current blocking is achievable by the wider bandgap interlayer. Hexagonal LuFeO₃ and its multiferroic variants (including, for example, LuFeO₃/LuFe₂O₄ superlattices) must be grown on GaN under conditions that do not chemically modify (or oxidize) the GaN or intervening Al(Ga)N layers.

### C. Polarization and Magnetization in Multiferroics

The multiferroic behavior of BiFeO₃ is partially governed by the stereochemically active lone pair electrons in the Bi 6s orbital, and their interaction with spins from the magnetic Fe atoms. The multiferroic behavior of LuFeO₃-LuFe₂O₄ superlattices is partially governed by a trimer distortion (e.g a physical rumpling of the $LuO_{1.5}$ layer involving a 2-up, 1-down positioning of Lu atoms) [35] which interacts with the spin moments of the neighboring iron ions. The difference in mechanisms causes the ferroelectric and magnetic transition temperatures to be largely different in each material. Accordingly, the strength of the magnetic and electrical-dipole moments is different for each. For SOT devices, it may be suitable to have a semiconductor channel carrier concentration similar to the charge concentration arising from the spontaneous polarization in the multiferroic material. For this reason, $LuFeO_3$ and $LuFeO_3$-$LuFe_2O_4$ superlattices are attractive, as well as La-substituted $BiFeO_3$ ($La_xBi_{1-x}FeO_3$) which reduces the spontaneous polarization compared to unsubstituted $BiFeO_3$ [39]. In theory, a SOT structure could be realized with only the magnetism from the multiferroic, but the relatively weak canted moments of $BiFeO_3$ and $LuFeO_3$-$LuFe_2O_4$ superlattices on the order of 0.05 μB/formula unit may necessitate coupling with a magnetic material with a larger moment. This may aid in the switching the electrical polarization by switching the magnetic moment of the multiferroic material.

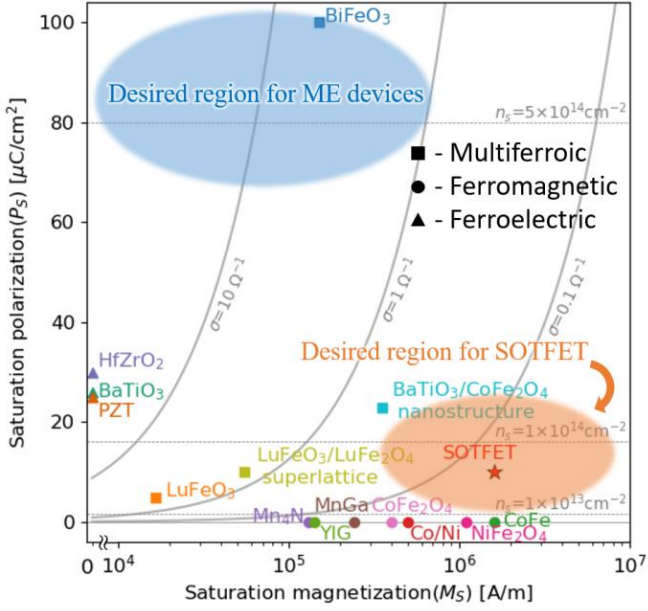

Fig. 3. Polarization and magnetization of various ferroelectrics [40], [41] (triangles); ferromagnets [29], [34], [42]-[47] (circles); and multiferroics [16]-[21], [34], [41], [48], [49] (squares). The solid gray curves represent fixed magnetoelectric transconductance, given by $\sigma = P_s/(\mu_0 M_s)$. The red star indicates the parameters used by [13] to theoretically demonstrate multiferroic switching in a SOTFET.

Figure 3 shows the saturation electrical polarization $P_s$ versus saturation magnetization $M_s$ for multiferroic materials and includes ferroelectric and ferromagnetic materials for comparison. The theoretical model of the SOTFET [13] suggests that switching a multiferroic's electrical polarization by first switching its magnetization is facilitated when the polarization is reduced and/or magnetization is increased (to enable strong exchange coupling to the magnetic layer). Therefore, we compare the relative strengths of polarization and magnetization in a multiferroic by introducing a parameter $\sigma = P_s/(\mu_0 M_s)$, which has units of conductance despite no conduction happening within the multiferroic. Fig. 3 utilizes σ, shown by the gray curves, as a figure of merit where low σ for a given magnetoelectric coupling energy is desired for the multiferroic in a SOTFET. We note, however, that a more appropriate figure of merit must be a more specific quantification of magnetoelectric coupling energy that measures how easily the polarization of the multiferroic can be driven by its magnetization. Nonetheless, the qualitative trend suggested by [13] supports our aim for low σ, and it can be used as a rule of thumb when engineering multiferroic properties for the SOTFET. In Fig. 3, we imply that a region with low σ is beneficial for the SOTFET where magnetization is switched first, while a region with high σ is advantageous for magnetoelectric devices that switch polarization first. This would necessitate efforts to increase magnetization and exchange coupling, such as epitaxially stabilizing a hidden ground state of a frustrated ferrimagnet like $LuFe_2O_4$ [21], [29] and/or decrease polarization, such as by substituting La or other rare-earth ions into $BiFeO_3$ [39], [50], [51]. Decreasing polarization in the multiferroic remains a priority because the expected polarization needed to control the semiconducting channel is on the order of 1 μC/cm², which is lower than most ferroelectrics and multiferroics, as shown in Fig. 3.

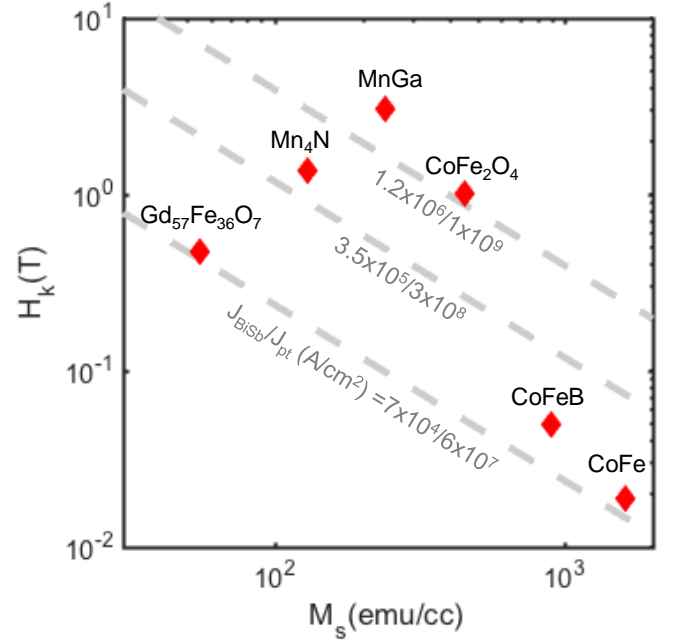

Fig. 4. Anisotropy field and saturation magnetization for various PMA magnets [16], [34], [54]-[61]. The gray dashed lines represent calculated, fixed critical switching currents through Pt and BiSb.

## IV. FERRO/FERRIMAGNETS

For energy efficient switching of a ferro/ferrimagnet using spin transfer torque (STT) or spin orbit torque (SOT), ferro/ferrimagnets with perpendicular magnetic anisotropy (PMA) are generally favored for scaling. This is because PMA magnets require lower switching currents than in-plane anisotropic magnets for a given energy barrier for switching [52]. While lowering the energy barrier leads to lower switching currents, practical magnetic memory applications require energy barriers greater than 40 $k_BT$ so that retention time against thermal agitation is at least 10 years [52]. Therefore, we limit our discussion of ferro/ferrimagnetic materials to only those that have been demonstrated to have PMA. However, we note that in-plane magnetized magnets have been shown to couple to $BiFeO_3$ [16], indicating that SOTFETs can be realized with magnetic materials that do not exhibit PMA.

The zero-temperature critical switching current of PMA magnets under single domain assumption can be written as $J_{C,perp}^{SH} = \frac{2e}{\hbar} \frac{M_s t_F}{\theta_{SH}} \left( \frac{H_{K,eff}}{2} - \frac{H_x}{\sqrt{2}} \right)$, where $H_{K,eff}$ is the effective anisotropy field representing the strength of the magnetic anisotropy, $H_x$ is an in-plane external magnetic field, $M_s$ is the saturation magnetization, $t_F$ is the thickness of the magnet, and $\theta_{SH}$ is the spin Hall angle of an adjacent spin-orbit material [53]. Fig. 4 shows the anisotropy field and saturation magnetization for several magnets scaled down to 1 nm thicknesses. The gray dashed lines show the critical switching current, calculated using the above equation, for several PMA magnets of 1 nm thickness, with the heavy metal Pt or the topological insulator BiSb as the spin orbit torque layer in the limit of negligible external in-plane assist field, which gives an upper bound for the switching current. The numbers in this plot are based on the anisotropy fields and saturation magnetizations of these materials.

In real cases, SOT switching may proceed through domain wall motion instead of the coherent reversal of a single domain, as has been observed by magneto-optical Kerr effect (MOKE) microscopy in W/CoFeB/MgO heterostructures [62]. SOT switching can also be thermally excited. Therefore, the dashed lines in Fig. 4 are the upper bounds of critical switching currents. For a SOTFET where the ferro/ferrimagnet and multiferroic magnetizations are strongly coupled, magnets with higher saturation magnetization may be advantageous because they raise the effective magnetization of the multiferroic, lowering its effective σ and facilitating the switching of its polarization. However, magnets with higher saturation magnetization are also more difficult to switch by SOT. The conflicting requirements indicate that the saturation magnetization of the magnet should be neither too high, nor too low for use in the SOTFET.

On the other hand, ferro/ferrimagnets with a lower anisotropy field have smaller energy barriers for switching. However, if the anisotropy field is too small, the magnet is thermally unstable. Therefore, the optimal ferro/ferrimagnets in Fig. 4 fall in an intermediate region for integration in the SOTFET. This parameter space must be further explored, however, given the range of competing effects. It is important to note that the critical switching current shown in Fig. 4 is assumed to flow in the spin-orbit layer only, not the ferro/ferrimagnetic layer. Current shunting through metallic ferro/ferrimagnets increases the actual total current required for SOT switching. From this point of view, insulating magnets, most of which are ferrimagnets such as yttrium iron garnet (YIG), $CoFe_2O_4$ and

$NiFe_2O_4$, are quite promising for energy efficient SOT devices. Furthermore, insulating magnets may aid the multiferroic as a gate dielectric in the SOTFET if the multiferroic layer has too small of a band gap and/or is too electrically leaky.

## V. SPIN-ORBIT MATERIALS

The spin-orbit materials work in conjunction with the ferromagnetic materials to determine the efficiency of magnetic switching by SOT. While properties of the ferromagnetic and multiferroic layers determine the energy barrier required to switch magnetization, the properties of the spin-orbit layer determines how much energy can be used to switch the magnet for a given current. Hence, we desire a spin-orbit material that can transfer the most spin angular momentum to the ferromagnet for a fixed amount of input voltage. A useful metric to quantify this efficiency is the spin Hall conductivity (SHC), which is the product of spin Hall angle and electrical conductivity. The spin Hall angle is given by $\theta_{SH} = 2e/\hbar (J_s/J_c)$, where $J_s$ is the spin current and $J_c$ is the charge current. In essence, the spin Hall angle evaluates how much spin angular momentum can be generated per unit of charge current density. Higher spin Hall angles are desired. High electronic conductivity in the spin-orbit layer is also desired in order to minimize current shunting into the ferromagnet if the ferromagnet is also electrically conductive. Fig. 5 shows the spin Hall angle and electronic conductivity for several potential spin-orbit materials. The gray dashed lines represent fixed SHCs, and the higher the SHC (the closer to the top-right corner of Fig. 5), the more desirable the spin-orbit material.

The most promising spin-orbit materials thus far have mainly been heavy metals, or topological insulators. The high nuclear charge of the atoms in heavy metals result in large spin-orbit coupling, leading to efficient charge-to-spin conversion. Heavy metals are also highly electrically conductive. Topological insulators are electrically insulating in the bulk, but have conductive surface bands. The spins of electrons that occupy the surface bands are locked to their momentum. This spin-momentum locking leads to very efficient charge-to-spin conversion. On the other hand, most of the charge current and spin-polarization is confined to the surfaces, leading to low electronic conductivity of topological insulators [68], [71]. Materials with both large spin Hall angle and high electronic conductivity, which fall in the top right corner of Fig. 5 are therefore highly desired for most efficient SOT switching in general, and for the SOTFET in particular.

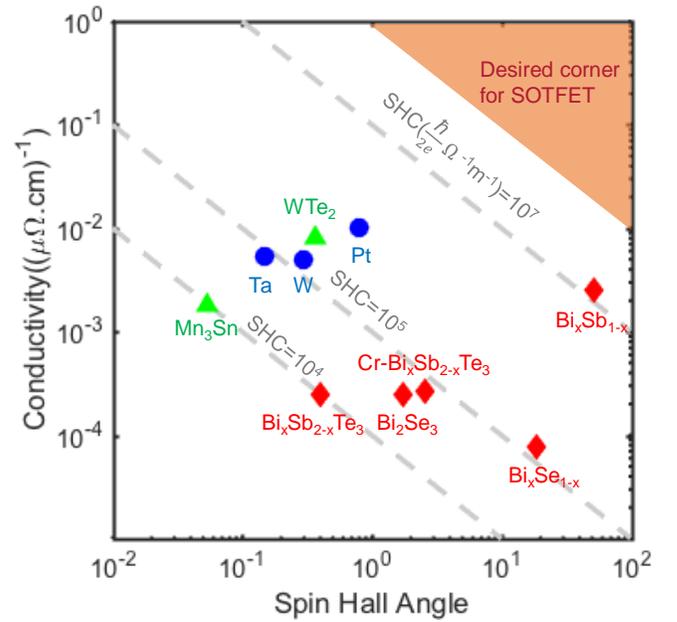

Fig. 5. Conductivity versus spin Hall angle for spin-orbit materials. Blue circles are heavy metals [61], [63], [64]; red diamonds are topological insulators [43], [65]-[69]; and green triangles are topological semimetals [70], [71]. The gray lines represent fixed spin Hall conductivity (SHC), which is the product of spin Hall angle and conductivity.

Generally, heavy metals have higher electronic conductivity, while topological insulators have higher spin Hall angles. To date, the highest spin Hall conductivity measured at room temperature was for BiSb, a conductive topological insulator. It exhibited a giant spin Hall angle with electronic conductivity comparable to a heavy metal [43]. As alternatives to heavy metals and topological insulators, Dirac and Weyl topological semimetals have recently been proposed as spin-orbit materials [72]. Dirac and Weyl topological semimetals are new states of topological quantum matter with a linear dispersion at the Dirac or Weyl points. Very recently, it has been found that there exists very large field-like torque at low temperature [72] as well as out of plane anti-damping torque through control of crystal symmetry in the semimetal WTe$_2$ [73], which offers probability of realizing deterministic SOT switching without the need for in-plane assisting magnetic fields. Thus, Fig. 5 offers a range of materials choices for the SO layer for the SOTFET. Compatibility with the underlying FM/MF/SC stack will determine which are practical.

## VI. Epitaxial Growth

While we have discussed the materials parameters that are desirable for each material type in the SOTFET, we also must consider how these materials can be compatibly integrated into a single structure. Several techniques may be used for the fabrication of the SOTFET, but epitaxy holds several potential advantages in the efficacy of each SOTFET material layer. High quality commercial semiconductor technology is epitaxially grown on semiconductor substrates, and the high degree of structural integrity of the crystals of epitaxial multiferroics assist in retaining their desired properties. Therefore, epitaxial growth will likely be required for the semiconductor and multiferroic layers in the SOTFET. Furthermore, the most efficient room temperature SOT switching to date had been demonstrated on an epitaxial structure [43], suggesting that it is important to explore and consider the epitaxial growth, ideally all in-situ, of the entire SOTFET structure to prove the feasibility of the device.

In epitaxial growth of thin films, it is desirable to generate smooth, single-crystalline films with low densities of defects such as -dislocations, stacking faults, vacancies, etc. One way to aid this is to utilize a substrate or template layer with the same or a very similar lattice constant, and a similar lattice symmetry as the desired film. This allows for a film to be grown in a very low strain state, which can prohibit the emergence of defects in the film that would act to relax the strain in the film. To this end, Fig. 6 plots the in-plane lattice constants of candidate SOTFET materials, as well as their energy band gaps. Energy band alignment is also important to consider since we want the multiferroic (and the ferro/ferrimagnetic layer if insulating) to act as a gate dielectric and the spin-orbit layer (and the ferro/ferrimagnetic layer if conductive) to act as a gate metal. Lattice matching of (001) BiFeO$_3$-based and (0001) LuFeO$_3$-based SOTFETs is shown via the dashed vertical lines in Fig. 6, indicating the possibility of epitaxially integrating several families of materials from topological insulators to semiconductors.

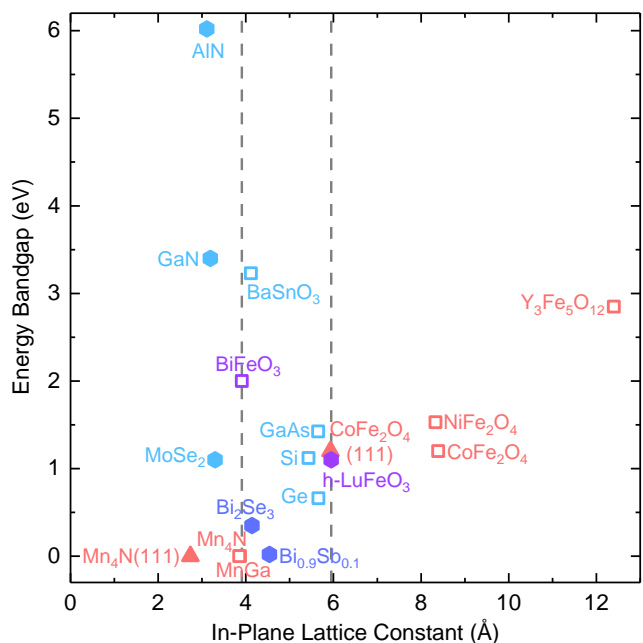

Fig. 6. Lattice constants and bandgaps for semiconductors [74]-[76] (teal); multiferroics [33], [38] (purple); ferromagnets [77]-[81] (orange); and topological insulators [82], [83] (blue). Hexagons represent hexagonal crystals, hollow squares represent cubic crystals, and triangles represent cubic crystals grown in the (111) direction. Materials with unlisted growth face planes are in the (001) orientation for the cubic crystals and in the (0001) orientation for the hexagonal crystals. The gray dotted lines are to guide the eyes along the lattice matching for SOTFETs based on BiFeO$_3$ and h-LuFeO$_3$.

By considering lattice constants, we can also utilize epitaxial strain to modify the material properties of each layer in the SOTFET. An example of epitaxial strain and symmetry sensitive material properties is that CoFe$_2$O$_4$, when grown on a hexagonal substrate, will adopt the (111) orientation to have proper symmetry matching, and when grown on a cubic substrate with (001) orientation, CoFe$_2$O$_4$ tends to adopt the (001) orientation and exhibit PMA [84]. When tensile strained, CoFe$_2$O$_4$ exhibits out-of plane anisotropy, and when compressively strained displays in-plane anisotropy [85]. Accordingly, material properties are tunable through epitaxial engineering.

## VII. Conclusion

SOTFET based structures are posed to yield a promising combination of spin-orbit torque phenomena with field effect control of electrons in semiconductors. This is expected to enable novel devices and architectures with combined memory and logic functionality. As high-quality, advanced materials such as topological insulators, ferrimagnetic insulators, and magnetoelectric multiferroics continue to be uncovered via epitaxial engineering, the rich physics that govern these phenomena will likely soon be elucidated as well.

### ACKNOWLEDGMENT

The research was partially supported by the National Science Foundation under Grant Nos. E2CDA 1740286 and NewLAW EFRI 1741694 and partially supported by the Semiconductor Research Corporation as nCORE task 2758. P.D.'s support by the National Science Foundation Graduate Research Fellowship under Grant No. DGE-1650441 is acknowledged.